\documentclass[pre,superscriptaddress,amsmath,amsfonts]{revtex4}
\usepackage[dvips]{graphicx}
\usepackage{color}
\begin{document}

\title{Ensemble equivalence for distinguishable particles}
\author{A. Fernandez-Peralta}
\affiliation{IFISC, Instituto de F{\'\i}sica Interdisciplinar y Sistemas Complejos (CSIC-UIB), Campus UIB, Palma de Mallorca, Spain}
\author{Raul Toral}
\affiliation{IFISC, Instituto de F{\'\i}sica Interdisciplinar y Sistemas Complejos (CSIC-UIB), Campus UIB, Palma de Mallorca, Spain}

\date{\today}

\begin{abstract}
Statistics of distinguishable particles has become relevant in systems of colloidal particles and in the context of applications of statistical mechanics to complex networks. When studying these type of systems with the standard textbook formalism, non-physical results such as non-extensive entropies are obtained. In this paper, we will show that the commonly used expression for the partition function of a system of distinguishable particles leads to huge fluctuations of the number of particles in the grand canonical ensemble and, consequently, to non-equivalence of statistical ensembles. We will see how a new proposed definition for the entropy of distinguishable particles by Swendsen [J. Stat. Phys. {\bfseries 107}, 1143 (2002)] solves the problem and restores ensemble equivalence. We also show that the new proposal for the partition function does not produce any inconsistency for a system of distinguishable localized particles, where the monoparticular partition function is not extensive.
\end{abstract}

\pacs{XXXX}

\maketitle

\section{Introduction}

The Gibbs paradox, namely, the entropy not being extensive for a classical ideal gas, is commonly resolved by adding an \emph{ad hoc} term to the entropy, $-k \log(N!)$ or, using Stirling formula, $-k N\log(N/e)$, where $N$ is the number of particles and $k$ Boltzmann's constant. This term appears if one divides the number of available states by $N!$, the so-called ``correct Boltzmann counting''. Its physical justification is usually attributed to quantum mechanics and the indistinguishable nature of identical particles. This line of reasoning has led to the believe that it is not possible to understand classically the origin of the $N!$ term~\citep{huang, pathria}. However, as early as 1921, Ehrenfest and Trkal~\citep{Ehrenfest} already argued that one must include the $N!$ term even for distinguishable particles (classical or non-identical quantum). The argument was retaken by van Kampen ~\citep{vanKampen} and some controversy on how we must regard statistics and entropy of distinguishable particles has arisen in the last years in a series of papers by Swendsen ~\citep{stat_mech_dist, Res_nag, entropy_works} and Nagle ~\citep{Nag1, Nag2}, amongst others ~\citep{Dieks, Cheng, Peters}.

This topic turns out not to be just of academic interest, but appears to be relevant in current research, where the importance of distinguishable statistics has increased to a large degree. For example, when studying colloids one has to take into account that no two colloidal particles are exactly alike, with differences in the mass and shape of each particle, and they are, consequently, distinguishable between them. In such a case, the indistinguishability explanation of the Gibbs paradox fails resoundingly and we have to consider other possible alternatives ~\citep{Colloids2, Colloids1} to explain experimental results with colloids. Another example is that of statistical mechanics of networks ~\citep{Newman, Bianconi, Bianconi2}, where edges/links of the network can be considered as being particles and pairs of vertices/nodes as energy states, establishing a straightforward analogy with quantum physical systems. In this case, links correspond to individual identifiable actions and it seems very forced to regard them as indistinguishable ~\citep{Oleguer1}. This distinguishability of links becomes important for multi-edged networks (where a pair of nodes can have more than one link) when performing statistics and entropy measures ~\citep{Oleguer2}.

In this paper we want to contribute to this topic by discussing the issue of ensemble equivalence. The problem appeared to us ~\citep{yo} when studying the grand canonical formalism of distinguishable particles (or network links). When using the common textbook expression of the partition function, anomalous fluctuations of the number of particles are obtained, leading to non-ensemble equivalence between the microcanonical/canonical and the grand canonical ensembles. The non-equivalence of ensembles casts doubts on the suitability of the statistical description. We will see how the inclusion of the $N!$ for distinguishable particles solves the problem and restores ensemble equivalence. Although it would seem that this inclusion should give non-extensive thermodynamic potentials for localized particles, we will show that this is not the case and the definition including the $N!$ term is consistent in all cases.

The paper is organized as follows: In section \ref{preliminary} we introduce the standard definitions of distinguishability, identical, etc., and other notation used in Statistical Mechanics, together with some simple Hamiltonian functions that will clarify those concepts. Section \ref{non-equivalence} presents the problem of ensemble equivalence for distinguishable particles, that arises when using the common textbook expression for the partition function and entropy. Finally, in section \ref{method} we explain how the definition of entropy introduced in~\cite{stat_mech_dist} solves this problem for all the examples given in section \ref{preliminary}. We end with some conclusions in section \ref{conclusions}.

\section{Preliminary concepts}
\label{preliminary}

{\bfseries Indistinguishable} particles are those described by a particle-exchange invariant set of microscopic states. If this invariance is not fulfilled, particles are called {\bfseries distinguishable}. 

In classical mechanics, a microstate for a system of $N$ particles corresponds to a point in phase space
\begin{equation}
\label{point_phase}
(q, p) \equiv (\vec{q}_1, \vec{q}_2, \dots, \vec{q}_N, \vec{p}_1, \vec{p}_2, \dots, \vec{p}_N),
\end{equation}
of generalized coordinates $\lbrace \vec{q}_{i} \rbrace$ and momenta $\lbrace \vec{p}_{i} \rbrace$. If the generalized coordinates and momenta of any two particles are exchanged, say $1$ and $2$
\begin{equation}
\label{point_phase_exchanged}
(\vec{q}_1, \vec{q}_2, \dots, \vec{q}_N, \vec{p}_1, \vec{p}_2, \dots, \vec{p}_N) \hspace{0.5cm} \rightarrow \hspace{0.5cm} (\vec{q}_2, \vec{q}_1, \dots, \vec{q}_N, \vec{p}_2, \vec{p}_1, \dots, \vec{p}_N),
\end{equation}
we obtain a different point in phase space (except for a set of points of zero measure). Consequently, in classical mechanics particle-exchange invariance never holds and particles are always distinguishable. This is independent on whether particles are identical or not. {\bfseries Identical} particles are those whose Hamiltonian is invariant under the exchange of any two particles,
\begin{equation}
\label{Hamil_symm}
{\cal H} (\vec{q}_1, \vec{q}_2, \dots, \vec{q}_N, \vec{p}_1, \vec{p}_2, \dots, \vec{p}_N) = {\cal H} (\vec{q}_2, \vec{q}_1, \dots, \vec{q}_N, \vec{p}_2, \vec{p}_1, \dots, \vec{p}_N).
\end{equation}

This definition of identical particles also holds in quantum mechanics if we consider that the quantum Hamiltonian depends on operators associated to the generalized coordinates and momenta as well as on operators associated to internal degrees of freedom such as spin.

In quantum mechanics, the $N$ particle microstate is a vector in the Hilbert space $|\psi\rangle$ which, in position representation and not considering internal degrees of freedom, is represented by a complex function $\psi (\vec{r}_1, \vec{r}_2, \dots, \vec{r}_N)$. A state of $N$ identical particles must be symmetric ($+$, bosons) or antisymmetric ($-$, fermions) under particle exchange
\begin{equation}
\label{wave_func_exchange}
\psi (\vec{r}_2, \vec{r}_1, \dots, \vec{r}_N) = \pm \psi (\vec{r}_1, \vec{r}_2, \dots, \vec{r}_N).
\end{equation}
Furthermore, according to the quantum mechanics postulates, particles are indistinguishable if and only if they are identical, otherwise they are distinguishable. 

Particles are called non-interacting or {\bfseries ideal} when it is possible to define monoparticular Hamiltonians ${ h}^{(i)}(\vec{q}_i,\vec{p}_i)$, depending only on the generalized coordinates and momenta of one particle, such that the $N$-particle Hamiltonian is ${\cal H} = \sum_{i=1}^{N} {h}^{(i)}$. In the quantum case, the monoparticular Hamiltonian might include spin and other quantum internal degrees of freedom.

For quantum systems (independently on whether particles are ideal or not), it is possible to construct the state $|\psi\rangle$ as a linear superposition of the eigenfunctions $|l_i\rangle$ of monoparticular Hamiltonians ${ h}^{(i)}$, defined as ${h}^{(i)}|l_i\rangle=\epsilon_{l_i}|l_i\rangle$:

-For distinguishable particles any state can be written as an unrestricted linear combination of elements of the product basis $\vert l_1\rangle |l_2\rangle\dots|l_N\rangle\equiv\vert l_1, l_2, \dots, l_N \rangle$. For an ideal system it is:
\begin{eqnarray}
\label{microstate}
{\cal H} \vert l_1, l_2, \dots, l_N \rangle &=& (\epsilon_{l_1}+\dots+\epsilon_{l_N}) \vert l_1, l_2, \dots, l_N \rangle.
\end{eqnarray}

-For indistinguishable particles one must use instead the suitably symmetrized or antisymmetrized basis or, alternatively, the second-quantization basis $||n_0, n_1, n_2, \dots\rangle\hspace{-1.2pt}\rangle$ in terms of the occupation numbers $n_\ell$ of individual levels $\ell$. 
For an ideal system it is:
\begin{eqnarray}
\label{microstate-b}
{\cal H} ||n_0, n_1, \dots\rangle\hspace{-1.2pt}\rangle &=& (n_{0} \epsilon_{0}+ n_{1} \epsilon_{1}+\dots) ||n_0, n_1,\dots\rangle\hspace{-1.2pt}\rangle.
\end{eqnarray}
 
In the ideal case, we can define the one-particle partition function ${\cal Z}_1^{(i)}$ associated to particle $i$. In the context of classical mechanics the definition is
\begin{equation}
\label{class_part_func}
{\cal Z}_1^{(i)} = \int \frac{d \vec{q}_{i}d \vec{p}_{i} }{h^{f_i}}e ^{-\beta {h}^{(i)}(\vec{q}_{i}, \vec{p}_{i})},\hspace{20pt} \beta=1/kT,
\end{equation}
where $T$ is the temperature and $f_i$ the number of degrees of freedom of particle $i$. The quantum counterpart is
\begin{equation}
\label{quantum_part_func}
{\cal Z}_1^{(i)} = \sum_{l_i} e^{-\beta \epsilon_{l_i}}.
\end{equation}

Non-interacting particles are said to be {\bfseries non-localized} if the monoparticular partition function ${\cal Z}_{1}$ fulfills
\begin{equation}
\label{part_nonloc1}
{\cal Z}_1(V, T) = V f(T),
\end{equation}
and {\bfseries localized} if it depends only on temperature
\begin{equation}
\label{part_loc1}
{\cal Z}_1(V, T) = \phi(T).
\end{equation}
Intuitively, localized particles correspond to those for which the eigenfunctions of the monoparticular Hamiltonian are localized in space. Examples being the infinitely-confining harmonic or infinite square well potentials in a finite region.

In order to fix ideas and to understand the concepts and definitions presented above, we will categorize the particles of four different Hamiltonians, whose statistics will be considered later. We restrict ourselves to the simplest examples (non-interacting particles) addressed in common textbooks of statistical mechanics. The results can be generalized to interacting particles as the nature of the problem addressed in this paper does not concern interactions.

(i) In the first example, we consider a non-relativistic gas of non-interacting identical particles without any internal or rotational degrees of freedom and not subject to any external field. The Hamiltonian can be written as 
\begin{equation}
\label{Ideal_gas_ham1}
{\cal H}= \sum_{i=1}^{N} \frac{{\vec{p}_i}^{\hspace{0.1cm}2}}{2 m}.
\end{equation}
We will see later on that those kind of particles are non-localized. Classically, this is a system of identical, distinguishable, non-localized particles. The quantum version represents a system of identical, hence indistinguishable, non-localized particles.

(ii) The second example is the previous ideal gas but each particle having a different mass. This classifies the particles as non-localized, non-identical and, hence, distinguishable both in the classical and quantum version. The Hamiltonian is
\begin{equation}
\label{Ideal_gas_ham2}
{\cal H}= \sum_{i=1}^{N} \frac{{\vec{p}_i}^{\hspace{0.1cm}2}}{2 m_{i}}.
\end{equation}
It can be considered as a crude representation of a system of colloidal particles, each one with a different mass.

(iii) The third example is a set of harmonic oscillators, each one oscillating around a different position $\vec{a}_{i} $
\begin{equation}
\label{harm_osc}
{\cal H}= \sum_{i=1}^{N} \left[ \frac{{\vec{p}_i}^{\hspace{0.1cm}2}}{2 m} + \frac{m \omega^2}{2} (\vec{r}_{i}-\vec{a}_i)^2 \right].
\end{equation}
This constitutes a system of non-identical, distinguishable, localized particles, both in the quantum and classical cases.

(iv) The final example is the statistics of paramagnetism, where we have a set of localized particles with magnetic moments $\lbrace \vec{\mu}_{i} \rbrace$ in a magnetic field $\vec{B}$
\begin{equation}
\label{param}
{\cal H}= \sum_{i=1}^{N} \left[ -\vec{\mu}_{i} \cdot \vec{B} + {h}^{(i)}_{\text{loc}} \right].
\end{equation}
Here ${h}^{(i)}_{\text{loc}}$ is an infinitely-confining Hamiltonian which localizes the particles around particular points $\vec a_i$ in space. Particles are then non-identical and distinguishable, both quantum and classically, despite the fact that the magnetic part of the Hamiltonian is invariant under particle exchange.

\section{non-ensemble equivalence}
\label{non-equivalence}
In the canonical ensemble, thermodynamic properties follow from the calculation of the partition function. In the quantum case, this is defined as
\begin{equation}
{\cal Z}_{N}=\sum_{m}e^{-\beta E_m},
\end{equation}
where $|m\rangle$ are the eigenstates of the Hamiltonian and $E_m$ the energy eigenvalues. For classical systems, the definition is
\begin{equation}
\label{part_clas}
{\cal Z}_{N}=\int \dfrac{d\vec{q}_1d\vec{p}_1}{h^{f_1}}\cdots \dfrac{d\vec{q}_Nd\vec{p}_N}{h^{f_N}}e^{-\beta{\cal H}}.
\end{equation}

For ideal systems, the usual calculation~\citep{huang, pathria} for distinguishable particles writes the partition function as 
\begin{equation}
\label{dist}
{\cal Z}_{N} = \prod_{i=1}^{N} {\cal Z}^{(i)}_{1}.
\end{equation}
Classically, this expression follows straightforwardly from ${\cal H} = \sum_{i=1}^{N} {h}^{(i)}$ and, in the quantum case, from Eq.(\ref{microstate}).

In the classical case (but not in the quantum case), if particles are identical, this expression simplifies to 
\begin{equation}
\label{disti}
{\cal Z}_{N} = [{\cal Z}_{1}]^N.
\end{equation}
 
The extension of Eq.(\ref{dist}) to quantum identical particles is non-trivial. While the exact calculation depends on whether the particles are fermions or bosons, an approximate result of general validity is obtained by replacing the factorial of the occupation number by $n_i!\approx 1$~\citep{Balescu}. The validity of this approximation improves at high temperatures where the mean occupation number is small. This leads to:
\begin{equation}
\label{indist}
{\cal Z}_{N} \approx \frac{[{\cal Z}_1]^{N}}{N!}.
\end{equation}

Quite generally, the partition function might depend on volume $V$, (inverse) temperature $\beta$ and number of particles $N$. Its derivatives provide the internal energy $U$, entropy $S$, pressure $P$ and chemical potential $\mu$:
\begin{eqnarray}
U &=& - \left(\frac{\partial \log {\cal Z}_N}{\partial \beta}\right)_{N,V},\label{can_U}\\
S&=& k \left(\frac{\partial (T\log {\cal Z}_N)}{\partial T}\right)_{N,V},\label{can_S}\\
P &=& k T \left(\frac{\partial \log {\cal Z}_N}{\partial V}\right)_{\beta,N},\label{can_P}\\
\mu &=& - kT \left(\frac{\partial \log {\cal Z}_N}{\partial N}\right)_{\beta,V}.\label{can_mu}
\end{eqnarray}

In the grand canonical ensemble, the number of particles is allowed to fluctuate. The probability of finding $N$ particles is given by
\begin{equation}
p(N)=\dfrac{z^N {\cal Z}_{N}}{\Xi (z,V,\beta)},
\end{equation}
with a grand canonical partition function defined as
\begin{equation}
\label{grand_can}
\Xi (z,V,\beta)= \sum_{N=0}^{\infty} z^N {\cal Z}_{N},
\end{equation}
and $z=e^{\beta\mu}$. The average (observable) number of particles is 
\begin{equation}
\langle N \rangle =\sum_{N=0}^{\infty} Np(N).
\end{equation}

The derivatives of the grand canonical partition function are:
\begin{eqnarray}
U &=& - \left(\frac{\partial \log \Xi}{\partial \beta}\right)_{V,z},\label{grand_U}\\
S &=& k \left(\frac{\partial (T\log {\Xi})}{\partial T}\right)_{\mu,V},\label{grand_S}\\
P &=& k T \left(\frac{\partial \log \Xi}{\partial V}\right)_{\beta,z},\label{grand_P}\\
\langle N \rangle &=& z \left(\frac{\partial \log\Xi}{\partial z}\right)_{\beta,V}\label{grand_N}.
\end{eqnarray}
Ensemble equivalence tells us that we can use either the canonical or the grand canonical formalism and still obtain the same functional form for $U(N,V,T)$, $S(N,V,T)$, $P(N,V,T)$, $\mu(N,V,T)$, provided we identify $N$ with $\langle N\rangle$ and take the limit $N\to \infty$. The physical basis of this equivalence relies on the fact that the probability $p(N)$ is heavily concentrated around its mean value $\langle N \rangle$ and hence this number can be identified as the physically measurable number of particles $N$. If this does not occur, ensemble equivalence is not justified and the whole building of Statistical Mechanics falls apart. A necessary condition for ensemble equivalence is that the fluctuations of the number of particles in the grand canonical ensemble, measured by the root-mean-square $\sigma[N]$, are vanishingly small compared to the average value $\langle N\rangle$. This usually requires the thermodynamic limit: $\lim_{\langle N\rangle\to\infty}\dfrac{\sigma[N]}{\langle N\rangle}=0$.

We will now show that ensemble equivalence does not hold if one uses the partition function given by Eq.(\ref{dist}) or Eq.(\ref{disti}), thus showing that those partition functions are not acceptable.

Let us take, for example, Eq.(\ref{disti}). The grand canonical partition function reads
\begin{equation}
\label{grand_can_dist}
\Xi = \sum_{N=0}^{\infty} z^N {\cal Z}_1^{N} = \frac{1}{1-z {\cal Z}_{1}}.
\end{equation}
The geometric sum is convergent as, using Eq.(\ref{grand_N}) we obtain
\begin{equation}
 z {\cal Z}_{1} = \frac{\langle N \rangle}{\langle N \rangle + 1} < 1.
\end{equation}
Indeed, using both sets of Eqs.(\ref{can_U}-\ref{can_mu}) or Eqs.(\ref{grand_U}-\ref{grand_N}) we obtain equivalent expressions (identifying $\langle N\rangle$ and $N$ and taking $N\to\infty$):
\begin{eqnarray}
U &=& -N \left(\frac{\partial \log {\cal Z}_1}{\partial \beta}\right)_{V},\label{thermodynamic_dist_U}\\
S & = & Nk \left(\frac{\partial (T \log {\cal Z}_1)}{\partial T}\right)_{V},\label{thermodynamic_dist_S}\\
P &=& k T N \left(\frac{\partial \log {\cal Z}_1}{\partial V}\right)_{\beta},\label{thermodynamic_dist_P}\\
\mu &=& - k T \log {\cal Z}_{1}.\label{thermodynamic_dist_mu}
\end{eqnarray}
However, when calculating the probability of the number of particles $N$ in the grand canonical ensemble, we obtain a geometric distribution with mean value and fluctuations:
\begin{eqnarray}
\label{N_distribution_dist}
p(N) &=& (z {\cal Z}_{1})^N (1-z {\cal Z}_{1}),\\
\langle N \rangle &=&\dfrac{z {\cal Z}_{1}}{1-z{\cal Z}_{1}},\\
\frac{\sigma[N]}{\langle N \rangle} &=& \sqrt{\frac{\langle N \rangle + 1}{\langle N \rangle}}\xrightarrow[{\langle N\rangle \to\infty}]{}1.
\end{eqnarray}
These huge fluctuations, as big as the average value, imply that the number of particles of the system is not well defined and, according to our discussion above, the partition function Eq.(\ref{dist}) is incorrect and cannot represent a macroscopic state of a physical system.

On the other hand, if one redoes these calculations with the partition function for indistinguishable particles Eq.(\ref{indist}), we still obtain equality of results, with the same functional forms Eqs.(\ref{thermodynamic_dist_U},\ref{thermodynamic_dist_P}) for the internal energy $U$ and pressure $P$, and a chemical potential $\mu = - k T \log( {\cal Z}_1/N)$ and entropy $S=Nk \left(\frac{\partial (T \log ({\cal Z}_1e/N))}{\partial T}\right)_{V}$. Furthermore, we obtain a Poisson distribution for the number of particles, with mean value and fluctuations:
\begin{eqnarray}
\label{N_distribution_indist}
p(N) &=& e^{-z {\cal Z}_1} \frac{(z {\cal Z}_1)^N}{N!},\\
\langle N \rangle &=& z {\cal Z}_1,\\
\frac{\sigma[N]}{\langle N \rangle} &=& \langle N \rangle^{-1/2}\xrightarrow[{\langle N \rangle \to\infty}]{} 0.
\end{eqnarray}
Note that the relative fluctuations now vanish in the thermodynamic limit. Then, apparently, if we follow the usual textbook procedure we obtain ensemble equivalence only for indistinguishable particles.

The validity of Eq.(\ref{dist}) for classical non-localized particles (e.g. an ideal gas) where ${\cal Z}_1\propto V$, is typically questioned because of the absence of extensiveness of the entropy Eq.(\ref{thermodynamic_dist_S}), or intensiveness of the chemical potential, Eq.(\ref{thermodynamic_dist_mu}), and a term $1/N!$ is added \emph{ad hoc} to the partition function Eq.(\ref{part_clas}) or by arguing that quantum indistinguishability resolves the Gibbs paradox. The failure of this explanation is that it implies that for quantum non-identical, and consequently distinguishable, particles the correct partition function should be Eq.(\ref{dist}), leading to an entropy which is not extensive. Here, we have shown that Eq.(\ref{dist}) is incorrect for all cases as it leads to non-ensemble equivalence, an unacceptable result from a statistical point of view. The question now is, which partition function is correct for each classification of particles?

\section{Correct partition function}
\label{method}
The answer to the previous question concerns very fundamental aspects of statistical mechanics, from the definition of entropy to the selection of a macrostate that only captures measures of macroscopic variables and we refer to \citep{stat_mech_dist, Res_nag, entropy_works} for a detailed discussion. The standard statistical expression of entropy is derived by considering one isolated system of $N$ particles and internal energy $E$ formed by two subsystems, $1$ and $2$, that can exchange energy and particles amongst them. The equilibrium condition is postulated to be that of a maximum for the probability for system $1$ to have energy $E_1$ and number of particles $N_1$. In order to recover the usual thermodynamic conditions for equilibrium, namely, equality of temperature, pressure and chemical potential, the entropy is defined (quantum) as
\begin{eqnarray}
\label{entropy_indist}
S &=& k \log \Omega, \hspace{20pt} \Omega = \sum_{m \left| \substack{E_{m}=E\\ N_{m}=N}\right.} 1,
\end{eqnarray}
where $|m\rangle$ is a microstate, eigenfunction of the Hamiltonian and the number operator with energy $E_m$ and number of particles $N_m$. $\Omega(E,N)$ is the total number of those microstates with $N$ particles and energy equal to $E$. In classical mechanics, the definition of entropy replaces the sum over microstates by an integral over phase space. The partition function for the canonical ${\cal Z}_{N}$ and grand canonical $\Xi$ ensembles follow from the definition of entropy~\citep{pathria}:
\begin{eqnarray}
\label{z_indist}
{\cal Z}_{N} &=& \sum_{m \left| N_{m} = N \right.} e^{-\beta E_{m}}, \\
\label{xi_indist}\Xi &=& \sum_{N=0}^{\infty} z^{N} {\cal Z}_{N}.
\end{eqnarray}

However, Swendsen~\citep{stat_mech_dist, entropy_works} claims that expressions Eq.(\ref{entropy_indist}), and consequently Eq.(\ref{z_indist}), are incorrect for distinguishable particles and that a new expression for the entropy is needed. He arguments, correctly in our opinion, that when writing the probability for system $1$ to have energy $E_1$ and number of particles $N_1$ and equivalent $E_2, N_2$ for system 2, one has to consider a multiplying factor $\binom{N}{N_1} = \frac{N!}{N_1! N_2!}$, that counts the number of ways of selecting $N_1$ distinguishable particles out of $N=N_1+N_2$. In simple words, the macrostate specifies that you have $N_1$ distinguishable particles in the system but, since particles can be exchanged, it is not possible to know the identity of the particles. Consequently, the correct relation between statistics and thermodynamics for distinguishable quantum particles is
\begin{eqnarray}
\label{entropy_dist}
S &=& k \log \left[ \frac{\Omega}{N!} \right], \hspace{20pt} \Omega = \sum_{m \left| \substack{E_{m}=E\\N_{m}=N}\right.} 1,\\
\label{entropy_dist_z}
 {\cal Z}_{N} &=& \frac{1}{N!} \sum_{m \left| N_{m} = N \right.} e^{-\beta E_{m}}, \hspace{20pt}\Xi = \sum_{N=0}^{\infty} z^{N} {\cal Z}_{N},
\end{eqnarray}
whereas for distinguishable (identical or not) classical particles the partition functions are:
\begin{eqnarray}
\label{entropy_dist2}
 {\cal Z}_{N} &=& \frac{1}{N!}\int \dfrac{d\vec{q}_1d\vec{p}_1}{h^{f_1}}\cdots \dfrac{d\vec{q}_Nd\vec{p}_N}{h^{f_N}}e^{-\beta{\cal H}}, \hspace{20pt}\Xi = \sum_{N=0}^{\infty} z^{N} {\cal Z}_{N}. 
\end{eqnarray}

We have shown in the previous section that when one applies Eqs.(\ref{z_indist}-\ref{xi_indist}) to distinguishable particles, non physical results are obtained, both from the thermodynamic (non-extensiveness of the entropy for non-localized particles) and statistical (non-vanishing fluctuations) points of view. A macrostate of distinguishable particles with expression given by Eq.(\ref{entropy_indist}) would imply that you know exactly which particles are forming the system, which is incompatible from the very definition of a macrostate, and this is exactly what the binomial coefficient $\binom{N}{N_1}$ is preventing.

Let us now work out the implications of the definitions Eqs.(\ref{entropy_dist}-\ref{entropy_dist2}) for each one of the Hamiltonians introduced in section \ref{preliminary}. 

\subsection{Ideal gas of identical non-localized particles}
Using Eq.(\ref{entropy_dist2}), the partition function of an ideal gas of classical distinguishable identical particles Eq.(\ref{Ideal_gas_ham1}) is
\begin{equation}
\label{part_ideal_gas_identical}
{\cal Z}_{N} = \frac{[{\cal Z}_1]^{N}}{N!}, \hspace{20pt} {\cal Z}_1 = V f(T), \hspace{20pt} f(T)=  \frac{(2 \pi m k T)^{3/2}}{h^3}.
\end{equation}
Note that the Gibbs paradox and huge fluctuations disappear immediately and classical distinguishability does not produce any incorrect prediction. 

\subsection{Ideal gas of non-identical non-localized particles}

Similar ideas can be extended to the ideal gas composed by non-identical particles. A simple extension of the previous arguments leads to the fact that, again, the presence of the $N!$ factorial term in the definition of the partition function is necessary if one wants a consistent statistical description. For the sake of concreteness, we consider a system in which particles are not identical because they have different masses, e.g. a system described by the Hamiltonian Eq.(\ref{Ideal_gas_ham2}). As the number of particles $N$ is macroscopic it is not feasible to specify the masses of each and everyone of the particles. Instead, we introduce a probability density function (pdf) $\rho(m)$ constructed from the histogram of all masses $ \lbrace m \rbrace\equiv(m_1,\dots, m_N)$ in the system.

In the statistical-mechanics derivation of entropy we consider a situation in which particles can be exchanged between systems $1$ and $2$. We use the label $n=1,\dots,\binom{N}{N_1}$ for each one of the different choices for $N_1$ particles in system $1$ and $N_2$ in $2$, and call $\lbrace m \rbrace_{1}^{n}=(m_{i_1},\dots,m_{i_{N_1}})$, $\lbrace m \rbrace_{2}^{n}=(m_{i_{N_1+1}},\dots,m_{i_{N}})$ the corresponding list of masses taken from $ \lbrace m \rbrace=(m_1,\dots, m_N)$. Again, as both $N_1$ and $N_2$ are macroscopic, we construct the pdf's $\rho^{n}_1(m)$ and $\rho^{n}_2(m)$ from the lists $\lbrace m \rbrace_{1}^{n}$ and $\lbrace m \rbrace_{2}^{n}$, respectively. The probability for system $1$ to have energy $E_1$ and number of particles $N_1$ is
\begin{equation}
\label{prob_mass}
p_{1}(E_1, N_1) = \sum_{n=1}^{\binom{N}{N_1}} \frac{\Omega_{1}(E_1, N_1, \rho_{1}^{n}(m))\Omega_{2}(E_2, N_2, \rho_{2}^{n}(m))}{\Omega(E, N, \rho(m))}.
\end{equation}

For large $N_1,\,N_2,\,N$ it is legitimate to assume that a vast majority of combinations $n$ will lead to the same distributions $\rho^{n}_1(m) = \rho^{n}_2(m) = \rho (m)$ and all the terms of sum Eq.(\ref{prob_mass}) contribute equally:
\begin{eqnarray}
p_{1}(E_1, N_1) &=& \binom{N}{N_1} \frac{\Omega_{1}(E_1, N_1, V_1, \rho(m))\Omega_{2}(E_2, N_2, V_2, \rho(m))}{\Omega(E, N, V, \rho(m))}\notag\\
&=&\dfrac{\Omega_{1}}{N_1!}\cdot\dfrac{\Omega_{2}}{N_2!}\cdot\dfrac{N!}{\Omega},\label{prob_mass_2}
\end{eqnarray}
leading to definitions Eqs.(\ref{entropy_dist}-\ref{entropy_dist_z}). This method can be generalized for whatever Hamiltonian of heterogeneous particles (such as colloids), that could depend on properties such as moment of inertia, shape of the molecules, etc. The correct partition function for non-identical distinguishable particles reads
\begin{equation}
\label{part_ideal_gas_identical2}
{\cal Z}_{N} = \frac{[{\cal Z}_1]^{N}}{N!}, \hspace{20pt} {\cal Z}_1 = V f(T), \hspace{20pt} f(T)= \frac{(2 \pi k T)^{3/2}}{h^3} \exp \left[ \int \rho(m) \log(m^{3/2}) dm \right],
\end{equation}
which has extensive entropy and vanishing fluctuations. Note here that Eq.(\ref{part_ideal_gas_identical2}) is exact for classical and quantum mechanics, and there is no need to resource to the Boltzmann approximation. If we choose $\rho(x) = \delta(x-m)$, we recover (\ref{part_ideal_gas_identical}) which is known to be exact for classical particles, but only an approximation for quantum systems. The limit of identical quantum particles is a singular one. By this we mean that, for quantum mechanics there is a discontinuity in considering the masses of the particles identical or not, because the Hamiltonian of the system is invariant under particle exchange or not, there is not a continuous transition there. 

Special attention deserves the case of a system with two (or more) macroscopically observable types of particles, $A$ and $B$. This is represented by a bimodal distribution of masses:
\begin{equation}
\label{bi_dist}
\rho(m) = \frac{N_A}{N} \rho_{A}(m) + \frac{N_B}{N} \rho_{B}(m), \hspace{20pt}N_A+N_B=N,
\end{equation}
where $N_A$ is the number of particles of type $A$ and $N_B$ of type $B$. Here $\rho_A(m)$ and $\rho_B(m)$ are non-overlapping distributions, peaked, respectively, around masses $m_A$ and $m_B$. 

The probability Eq.(\ref{prob_mass_2}) and the partition function Eq.(\ref{part_ideal_gas_identical2}) describe a system with a constant proportion of particles $A$ and $B$. This means that when the number of particles $N_1$ in subsystem 1 is fixed, we automatically fix the number of particles of $A$ and $B$ in this subsystem as $N_{1A}= N_1\cdot\dfrac{N_A}{N}$, $N_{1B}= N_1\cdot\dfrac{N_B}{N}$. This is the situation for a  system that, macroscopically, is not allowed to exchange particles selectively, i.e. no selective membranes. If one is interested in a situation in which the number of particles $N_{1A}$ and $N_{1B}$ can change independently at the macroscopic level, it is necessary to consider a macrostate that specifies not only the total number of particles of the system $N_1$, but the number of each type $N_{1A}$, $N_{1B}$, keeping $N_{1A}+N_{1B}=N_1$, and calculate its probability $p_{1}(E_1, N_{1A}, N_{1B})$. Rewriting $\Omega_{1}(E_1, N_1, \rho_{1}^{n}(m)) = \Omega_{1}(E_1, N_{1A}, N_{1B}, \rho_{A}(m), \rho_{B}(m))$ (we again assume that each one of the possible combinations $n$ leads to a same distribution of masses for $A$ and $B$), and splitting the $\binom{N}{N_1}$ configurations according to the value of $N_{1A}$, $\binom{N}{N_1}=\sum_{N_{1A}=0}^{N_1} \binom{N_A}{N_{1A}} \binom{N_B}{N_{1B}}$, Eq.(\ref{prob_mass}) becomes
\begin{equation}
\label{prob_mass3}
p_{1}(E_1, N_1) = \sum_{N_{1A} = 0}^{N_1} \binom{N_A}{N_{1A}} \binom{N_B}{N_{1B}} \frac{\Omega_1 \Omega_2}{\Omega} = \sum_{N_{1A} = 0}^{N_1} p_{1}(E_1, N_{1A}, N_{1B}).
\end{equation}
Using the probability function $p_{1}(E_1, N_{1A}, N_{1B})$ to derive the appropriate expression for the entropy and the partition function, we obtain
\begin{eqnarray}
\label{part_ideal_gas_identical3}
{\cal Z}_{N_A, N_B} &=& \frac{[{\cal Z}_{1A}]^{N_A}}{N_A!} \frac{[{\cal Z}_{1B}]^{N_B}}{N_B!},
\notag\\
{\cal Z}_{1A} &=& V f_{A}(T), \hspace{20pt} f_A(T)= \frac{(2 \pi k T)^{3/2}}{h^3} \exp \left[ \int \rho_A(m) \log(m^{3/2}) dm \right],
\notag\\
{\cal Z}_{1B} &=& V f_{B}(T), \hspace{20pt} f_B(T)= \frac{(2 \pi k T)^{3/2}}{h^3} \exp \left[ \int \rho_B(m) \log(m^{3/2}) dm \right].
\end{eqnarray}
Note the presence of the terms $N_A!$, $N_B!$, which will ensure that fluctuations of the number of particles $N_A$ and $N_B$ in the grand canonical ensemble vanish in the thermodynamic limit.

\subsection{Localized particles}

It would appear that the new definition brings problems in the case of localized distinguishable particles since the entropy and other thermodynamic potentials derived from Eq.(\ref{entropy_dist_z}-\ref{entropy_dist2}) appear to be non-extensive, at least in the ideal case where ${\cal Z}_1\propto V^0$. We will now argue that this is not the case.

For localized particles, e.g. those described by the Hamiltonians Eq.(\ref{harm_osc}) and Eq.(\ref{param}), we face the same conceptual problem than before: a macrostate cannot include the detailed location of each and every particle $\lbrace \vec{a}_{i} \rbrace$. Let us assume that there exists a set $M\ge N$ of available locations. A microstate of the system can be written as linear combination of the basis
\begin{equation}
\label{micro_loc}
\vert m \rangle = \vert \ell_1, \vec{a}_{1} \rangle \vert \ell_2, \vec{a}_{2} \rangle \dots \vert \ell_N, \vec{a}_{N} \rangle,
\end{equation}
where $\vec{a}_{i}$ is the actual location of the particle for this microstate, and $\ell_{i}$ is its monoparticular level. The energy of each level can be obtained solving ${h}^{(i)} \vert \ell_{i} \rangle = \epsilon_{\ell_{i}} \vert \ell_{i} \rangle$
\begin{eqnarray}
\label{monoparticular_loc}
\epsilon_{\ell} = \left(\ell+\frac{1}{2}\right) \hbar \omega, \hspace{1cm} \ell &=& 0, 1, \dots \hspace{2.0cm} \text{(Harmonic oscillators 1d)},
\notag\\
\epsilon_{\ell} = - g \mu_{B} B \ell, \hspace{1cm} \ell &=& -J, -J+1, \dots, J \hspace{1cm} \text{(Paramagnetism)},
\end{eqnarray}
where $\mu_{B}$ is the Bohr magneton, $g$ is the Land{\'e} g-factor and $J$ is the total angular momentum of the particles and, in the paramagnetic case, we have neglected the contribution from $h_\textrm{loc}^{(i)}$ to the energy.
The macroscopic description can not specify the location of each particle. Therefore, when calculating the partition function Eq.(\ref{entropy_dist_z}), we have to consider the count over the ways $N$ distinguishable particles can be placed in $M$ locations, i.e. $ \frac{M!}{(M-N)!}$, obtaining
\begin{equation}
\label{part_loc}
{\cal Z}_{N} = \frac{1}{N!} \sum_{m \vert E_m=E} e^{-\beta E_{m}} = \frac{1}{N!} \frac{M!}{(M-N)!} [{\cal Z}_1]^{N} = \binom{M}{N} [{\cal Z}_1]^{N}.
\end{equation}
The monoparticular partition function for the cases of interest is
\begin{equation}
\label{monoparticular_part_loc}
{\cal Z}_{1} = \phi(T) = \left\{ 
 \begin{array}{lcl}
 \left[2 \sinh \left( \frac{\beta \hbar \omega}{2} \right) \right]^{-1} \hspace{2.9cm} \text{(Harmonic oscillators)},\\
 \sinh \left[ \left(1+\frac{1}{2 J} \right) x \right]/\sinh \left[ \frac{1}{2 J} x \right] \hspace{1.5cm} \text{(Paramagnetism)},
 \end{array} \right.
\end{equation}
where $x=\beta g \mu_B J B$. The partition function in the grand canonical ensemble is
\begin{equation}
\label{grand_part_loc}
\Xi = \sum_{N=0}^{M} z^{N} {\cal Z}_{N} = \sum_{N=0}^{M} \binom{M}{N} \left[ z {\cal Z}_{1} \right]^N = \left( 1 + z {\cal Z}_{1} \right)^{M}.
\end{equation}
We can now prove ensemble equivalence. Using the standard relations, both the canonical and grand canonical ensembles (identifying $N$ and $\langle N\rangle$) lead to
\begin{eqnarray}
\label{thermodynamic_loc}
U &=& -N \frac{\partial \log {\cal Z}_1}{\partial \beta},
\notag\\
S&=& Nk\left[\dfrac{\partial(T\log {\cal Z}_1)}{\partial T}+(1-1/\alpha)\log(1-\alpha)-\log \alpha\right],\notag\\
P &=& k T N \frac{\partial \log {\cal Z}_1}{\partial V},
\notag\\
\mu &=& - k T \log \left( \frac{1-\alpha}{\alpha} {\cal Z}_{1} \right),
\end{eqnarray}
where $\alpha = N/M$. Note that despite being ${\cal Z}_1\propto V^0$, the entropy is extensive while the chemical potential keeps its intensiveness if we consider that the number of available locations $M$ is itself extensive and thus $\alpha = O(1)$. 
The probability of the number of particles follows now a binomial distribution
\begin{eqnarray}
\label{N_distribution_loc}
p(N) &=& \binom{M}{N} \alpha^N \left(1-\alpha\right)^{M-N},
\notag\\ 
\frac{\sigma[N]}{\langle N \rangle} &=& \langle N \rangle^{-1/2} \sqrt{1-\alpha}\xrightarrow[{\langle N\rangle \to\infty}]{}0,
\end{eqnarray}
whose fluctuations vanish in the thermodynamic limit, restoring ensemble equivalence.

The particular case typically considered in textbooks is $N=M$, i.e. the number of available locations equal to the number of particles, brings no further problems as in this case, we simply recover Eqs.(\ref{thermodynamic_dist_U}-\ref{thermodynamic_dist_P}). Note, however, that in such a case, it is impossible for the system to include more particles and, consequently the chemical potential $\mu \rightarrow \infty$ as $\alpha \rightarrow 1$. In the same limit in the grand-canonical ensemble it is $p(N)\to \delta(N-M)$ and fluctuations become exactly zero $\sigma[N]/\langle N \rangle = 0$. 

\section{Conclusions}
\label{conclusions}
We have shown that the common textbook expression of the partition function of a system of distinguishable particles which does not include the $N!$ term, leads to abnormally large fluctuations of the number of particles in the grand canonical ensemble. This occurs independently on whether particles are classical or quantum, localized or non-localized. The large fluctuations go against the  postulates of statistical mechanics, which require that the relative fluctuations of the number of particles in the grand canonical ensemble vanish in the thermodynamic limit, such that it is possible to identify the mean value of the number of particles as the physically measurable $N$ and ensure ensemble equivalence.

As argued by Swendsen~\cite{stat_mech_dist}, the correct partition function (including the $N!$ term) for distinguishable particles is obtained from the statistical derivation of entropy and the selection of a macrostate that only captures macroscopic measurements. We have tested this expression of the partition function with some common examples of distinguishable particles that include: an ideal gas of classical identical particles, an ideal gas of classical or quantum non-identical particles (where each particle holds a different mass), a set of harmonic oscillators and the statistics of paramagnetism. For the ideal gas of non-identical particles, we have discussed the derivation of the correct partition function for a unimodal and bimodal distributions of mass. We have also shown that the new proposal for the partition function does not produce any inconsistency for a system of localized particles, where the monoparticular partition function is not extensive ${\cal Z}_{1} \propto V^{0}$. The thermodynamic potentials obtained for all the examples fulfill the corresponding extensiveness properties and the fluctuations of the number of particles in the grand canonical ensemble vanish in the thermodynamic limit, restoring ensemble equivalence.

\section*{Acknowledgements}
We acknowledge useful discussions with O. Sagarra and C.J. P{\'e}rez-Vicente concerning the issue of distinguishability in networks. This work was supported by FEDER (EU) and MINECO (Spain) under Grant INTENSE@COSYP (FIS2012-30634) and EC project LASAGNE FP7-2012-STREP-318132. A.F-P acknowledges support by the FPU program of MECD (Spain).

\bibliographystyle{unsrt}
\bibliography{biblio}

\end{document}